\documentclass[sigconf,natbib=true]{acmart}
\usepackage{tikz}
\usetikzlibrary{calc,arrows.meta,positioning}
\AtBeginDocument{%
  }

\setcopyright{acmlicensed}
\copyrightyear{2027}
\acmYear{2027}
\acmDOI{XXXXXXX.XXXXXXX}
\acmConference[CHIIR27]{Make sure to enter the correct
  conference title from your rights confirmation email}{March 7--11,
  2027}{Berlin, Germany}
\acmISBN{978-1-4503-XXXX-X/2018/06}

\begin{document}

\title{When Agentic Retrieval Fails Silently: Framework and Production Evidence from a Creative AI Assistant }


\author{Yan (Fergie) Liang }
\affiliation{%
  \institution{Adobe}
  \city{San Francisco}
  \state{CA}
  \country{USA}
}

\author{Anuj Agarwal
}
\affiliation{%
 \institution{Adobe}
  \city{Noida}
  \state{Uttar Pradesh}
  \country{India}}

\author{Aditya Reddy Cheruku}
\affiliation{%
  \institution{Adobe}
  \city{San Jose}
  \state{CA}
  \country{USA}}


\begin{abstract}
   CTR, MRR, and behavioral NDCG assume the user sees ranked results and acts on them. In an agentic assistant, a generation layer consumes retrieval output first, and that assumption breaks in three ways. The agent reads the results before the user does (who). The user’s action can come several turns or sessions after retrieval (when). Outcomes blend retrieved content with the model’s own knowledge (what caused it). We call this the attribution gap. We propose Layered Agentic Retrieval Assessment (LARA), which measures retrieval quality, engagement and outcome, and product success as three independent layers joined by a four-field lineage schema that requires no new modeling. We examine the motivation on Adobe Express, where one retrieval infrastructure serves a gallery search surface and an agentic assistant at roughly one-eighth its session volume for the same user population. Chat users reformulate more than gallery users (44.4\% vs. 33.1\%) because chat reformulation refines generated output.  LARA restores attribution for the retrieval.

\end{abstract}




\maketitle

\section{Introduction }
\label{sec:intro}

Click-through rate (CTR), mean reciprocal rank (MRR), and normalized discounted cumulative gain (NDCG) \cite{ref2} rest on three assumptions about the interaction: the user sees the ranked results, the user acts on them soon after the query, and what the user does next reflects retrieval quality. In an agentic AI assistant, search becomes a subagent whose output is consumed by a generation layer before the user sees anything. Each assumption fails, and each failure is a structural shift rather than a tuning problem.

\textbf{Shift 1: the agent reads the results first (who).} In gallery search, a user types “birthday invitation,” scans 40 templates, and clicks one; the click grades retrieval. In an assistant, the user asks for a birthday invitation, the search subagent retrieves templates, and the agent picks one and builds the design. The user never sees the other 39. There may be no click at all, and whatever the user does see has already been filtered and reordered by the agent.

\textbf{Shift 2: action and retrieval come apart in time (when).} In a conversation, the same asset can be retrieved in turn 1 and again in turn 3, then exported in turn 5 or in a different session. Without an identifier that links the action to a specific retrieval event, timing cannot say which retrieval earned it.

\textbf{Shift 3: retrieval and generation blend (what caused it).} A bad outcome cannot be split between retrieval failure and generation failure, and a good outcome can hide a retrieval failure the model quietly compensated for.

Without per-component instrumentation, teams rely on session-level proxies such as retry, abandonment, and task completion. Retry alone fails in both directions. It fires on generation failures unrelated to retrieval, and stays silent when the model covers for a retrieval failure. We call this the attribution gap. Prior work frames agentic retrieval \cite{ref8}, reintroduces fine-grained feedback into generative search \cite{ref9}, and documents technical debt in production ML \cite{ref5} and drift in agentic systems \cite{ref10}, but leaves retrieval attribution in production unresolved.

We make three contributions. (1) A who/when/why account of how agentic assistants break behavioral retrieval attribution, mapped to the metric failures each shift causes (\S\ref{ssec:to-fixes}). (2) LARA, three independent measurement layers joined by a four-field lineage schema that restores attribution without new modeling (\S\ref{ssec:layers}--\ref{ssec:disclosure}). (3) Production evidence from Adobe Express, where one retrieval infrastructure serves a gallery search surface and an agentic assistant for the same user population (August 2026), including a reformulation rate that runs opposite to the naive expectation and a retrieval rate of 0.31\% on assistant prompts (\S\ref{sec:evidence}--\ref{sec:unexpected}). LARA evaluates retrieval once a retrieval event fires; whether retrieval should fire at all is a planner-routing question we name Layer 0 and leave to future work.

\section{The LARA Framework }
\label{sec:framework}

\subsection{From shifts to failures to fixes }
\label{ssec:to-fixes}

Table \ref{tab:framework} is the core of the framework. Each shift breaks one assumption, makes a familiar metric misleading in a specific way, and is resolved by one LARA component. One failure sits partly outside any automated boundary: a confidently presented wrong answer that the user never checks. Layer 1 catches it only while offline relevance labels are current, so deployments should report it as a known blind spot rather than undercount it silently.

\begin{table*}[htb]
    \centering
    \begin{tabular}{p{.8in}p{.8in}p{1.5in}p{1.5in}p{1.1in}}
    \hline\hline
      \textbf{Shift}  & \textbf{Assumption broken}  & \textbf{What today’s metrics report}  & \textbf{Why it misleads}  & \textbf{LARA fix} \\\hline
\textbf{1. The agent reads first} (who)  & The user sees the results  & CTR = 0 with no click; CTR and MRR score only the agent-chosen subset; zero-click answers count as abandonment  & Clicks grade curation, not retrieval; unshown results go unevaluated; success can read as failure (good abandonment \cite{ref13})  & Layer 1 scored on the full ranked list at fire time (returned\_asset\_ids); outcomes reported by retrieval role \\\hline
\textbf{2. Action is delayed} (when)  & The user acts right after the query  & Exports and completion carry no link back to the supplying retrieval  & An export several turns after retrieval cannot be credited to it; exports landing in a new session read as abandonment  & consumed\_asset\_id and source\_retrieval\_id on every action event \\\hline
\textbf{3. Retrieval and generation blend} (what caused it) & The outcome reflects retrieval  & Retry or reformulation reads as retrieval failure; no retry reads as success  & Retry fires on generation failures; the model’s own knowledge can hide a stale or wrong retrieval  & Layer 1 measured independently of behavior, read against Layer 2 failure signals (\S\ref{ssec:reading}) \\\hline\hline
    \end{tabular}
    \caption{The three structural shifts, the metric failures each causes, and the LARA component that resolves it. }
    \label{tab:framework}
\end{table*}

\noindent Shift 3 is the most dangerous of the three, because nothing in the session-level signal distinguishes it from success (\S\ref{ssec:reading}).

\subsection{Three layers }
\label{ssec:layers}

LARA measures each pipeline stage at the earliest point where its failure becomes observable.

\begin{figure}[htb]
    \centering
    \begin{tikzpicture}[
        box/.style={draw, very thick, text width=2.3in, align=left, inner sep=6pt, font=\small, minimum height=0.5in},
      ]
      \node[box] (l3) {\textbf{LAYER 3: PRODUCT SUCCESS}\\[3pt] task completion rate \textperiodcentered\ goal fulfillment rate \textperiodcentered\ export and share rate};
      \node[box, fill=black!8, below=0pt of l3] (l2) {\textbf{LAYER 2: ENGAGEMENT AND OUTCOME}\\[3pt] GPR \textperiodcentered\ ACR \textperiodcentered\ ORF \textperiodcentered\ UOR};
      \node[box, below=0pt of l2] (l1) {\textbf{LAYER 1: RETRIEVAL QUALITY}\\[3pt] null rate \textperiodcentered\ precision@K \textperiodcentered\ NDCG@K (server-side logs)};
      \draw[dashed, thick] (l3.north east) rectangle ($(l1.south east)+(0.6in,0)$);
      \coordinate (arrowbot) at ($(l1.south east)+(0.12in,0.15in)$);
      \coordinate (arrowtop) at ($(l3.north east)+(0.12in,-0.15in)$);
      \draw[-{Latex[length=2.2mm]}, thick] (arrowbot) -- (arrowtop);
      \coordinate (arrowmid) at ($(arrowbot)!0.5!(arrowtop)$);
      \node[rotate=90, font=\small] at ($(arrowmid)+(0.33in,0)$) {lineage schema};
    \end{tikzpicture}
    \caption{LARA’s three layers. Layer 1 is computable from server-side logs alone. Layers 2 and 3 require the lineage schema from \S2.4. The schema (right) connects all three layers across event types and turns. }
    \label{fig:layers}
    \Description{LARA's three layers.}
\end{figure}
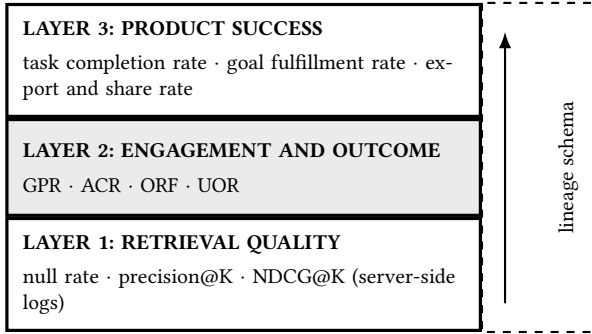

\textbf{Layer 1:} Retrieval quality. Did the search subagent return the right assets, in the right order? Metrics are null rate, precision@1, precision@K, and NDCG@K, computed from server-side logs against offline relevance labels (human or LLM-as-judge \cite{ref7}) at the moment retrieval fires, independent of anything the user or agent does next. CTR becomes a secondary directional signal. Layer 1 is the only layer that can detect a silent retrieval failure the model compensated for.

\textbf{Layer 2:} Engagement and outcome. Did anyone use what retrieval returned, and did anything signal that retrieval was wrong? Layer 2 keeps positive evidence and failure evidence separate and reports both by retrieval role, a tag written at retrieval time that records what retrieval was for: direct retrieval, retrieve-then-inspect, retrieve-then-act, reference input for generation, an embedded step in a larger task, or context-dependent.

\textit{Genuine Positive Rate (GPR)} is the share of retrievals followed by user-initiated use of a returned asset after the response is shown, either a follow-up prompt that continues with it or a direct interaction such as selecting, editing, or exporting it.

\textit{Agent Consumption Rate (ACR)} is the share of retrievals where the agent uses a returned asset inside the same request, before the user has a result to act on (e.g., finding a logo and removing its background because the user asked for both in one prompt). This is positive evidence that the result entered the workflow, and it is counted, not discarded. It is tracked separately from GPR because it carries different evidentiary weight: the agent using an asset is the agent’s judgment, while a user’s follow-up action is independent confirmation. A retrieval can register in both GPR and ACR if the user later acts on the same asset the agent already consumed.

\textit{Observed Retrieval Failure Rate (ORF)} is the share of retrievals followed by at least one of three retrieval-specific signals, Search Correction (the user says the retrieved object or route was wrong), Search Reformulation (the user tries again to retrieve for the same unresolved need), or Agentic Rejection (the workflow expected to consume a returned asset but used something else). A Downstream Correction, which fixes an edit, answer, or generation while keeping the retrieved asset, is not a retrieval failure. The Observed Non-Failure Rate (ONFR = 1 $-$ ORF) is reported alongside GPR and ACR and never read as satisfaction.

\textit{Unobserved Outcome Rate (UOR)} is the share of retrievals with no evidence of use by the user or the agent and no failure signal. It is reported as its own category, not as abandonment or as success.

\textbf{Layer 3:} Product success. Did the user accomplish the goal? In creative tools the clearest signals are export, share, and download, because the goal is a work artifact. Task completion is defined per retrieval role rather than globally, and goal fulfillment requires LLM-as-judge \cite{ref7} or human annotation. Layer 3 becomes a retrieval signal, not only a product-health signal, when source\_retrieval\_id links the success event to the retrieval that supplied the asset.

\subsection{Reading the layers together }
\label{ssec:reading}

Reading Layer 1 against Layer 2 locates the failing stage (Table \ref{tab:diagnosis}). The last row is the most dangerous. Without Layer 1 it is invisible; with Layer 1 it is detectable when retrieval fires, before complaints accumulate.

\begin{table}[htb]
    \centering
    \begin{tabular}{lp{1.2in}p{1.2in}}
    \hline\hline
      Layer 1  & Layer 2  & Diagnosis \\\hline
Low  & ORF high  & Retrieval returned the wrong assets \\
High  & ORF low; Downstream Corrections high  & Generation failed on good assets \\
High  & ORF high (Agentic Rejection)  & Agent bypassed good results (parametric blending) \\
Low  & ORF near zero  & Silent retrieval failure masked by the model\\\hline\hline
\end{tabular}
    \caption{Cross-layer diagnosis. }
    \label{tab:diagnosis}
\end{table}

\subsection{Lineage schema }

In current production logs, retrieval events and action events share no field that connects them. Four fields close the gap. retrieval\_id is a unique identifier on every retrieval event. returned\_asset\_ids is the full ranked list at fire time, before the generation layer filters or reorders it; alone it enables Layer 1 and is the highest-value single change. consumed\_asset\_id is the asset on every action event, whether a user action or an agent tool call. source\_retrieval\_id is the retrieval\_id that supplied that asset, on the same action event; for an asset already placed in the document, it is inherited from the placement event. All four are logging additions, not predictive models; no new modeling is required.

Because source\_retrieval\_id names a retrieval rather than a session, it resolves cross-turn recurrence (Shift 2) and survives session boundaries, so exports landing in a new session stay attributable. Joined, the four fields recover the retrieval rank of the asset actually used, independent of display order, letting RAG evaluation frameworks such as RAGAs \cite{ref3} and ARES \cite{ref4} use direct attribution instead of timing heuristics. The retrieval role tag is written on the retrieval event alongside them.

\subsection{Reporting disclosure }
\label{ssec:disclosure}

No shared standard yet covers production instrumentation of agentic retrieval, as TREC \cite{ref6} and TREC iKAT \cite{ref1} do for traditional and interactive IR. Deployments reporting LARA should disclose which lineage fields are implemented (or whether attribution is timing-based), how retrieval role is assigned, which actions count as user-initiated use, whether P95 latency and tokens per successful task are logged, and whether telemetry passed automated validation \cite{ref5}. Without this, Layer 2 and Layer 3 figures across deployments are not comparable.

\begin{table*}
    \begin{tabular}{llllll}
    \hline\hline
       Metric  & Gallery  & Agentic (all)  & Agentic chat  & Agentic editor  & $\Delta$ \\\hline
Median words per query  & 2  & 10  & 6  & 10  & 5× \\
Queries $\leq$ 2 words  & $\sim$71\%  & 11.6\%  & 27.3\%  & 10.7\%  & $-$59 pp \\
Reformulation rate  & 33.1\%  & 31.2\%  & 44.4\%  & 30.6\%  & +11 pp (chat) \\
Meaningful-action rate  & 40.5\% / 46.7\%  & 32.5\%  & 40.5\%  & 32.2\%  & $-$8 pp \\
Abandonment rate  & 53.3\%  & 67.5\%  & 59.5\%  & 67.8\%  & +14 pp \\
Prompts with retrieval  & 100\%  & 0.31\%  & 10.7\%*  & $\sim$0\%  & --- \\
\hline\hline
    \end{tabular}
    \caption{Legacy behavioral proxies, gallery search vs.\ agentic assistant (Adobe Express, August 2026; same retrieval infrastructure and user population). Session, user, and query counts are withheld per Adobe data-disclosure policy; gallery operates at roughly 8--9$\times$ the agentic assistant’s volume across all three. Gallery meaningful action shown at turn / session grain; agentic meaningful action = export or component interaction (select-variation, slider, color picker, grid click). Length medians are authoritative given an extreme outlier in the agentic distribution. *Directional; see \S\ref{sec:unexpected} on missed retrieval. }
    \label{tab:stats}
\end{table*}

\section{Production Evidence from Adobe Express }
\label{sec:evidence}

Adobe Express serves a gallery search surface and an agentic AI assistant, with chat and in-editor entry points, from the same retrieval infrastructure to the same user population. We analyzed production sessions from August 2026. Table \ref{tab:stats} reports the behavioral proxies teams compute today. None of LARA’s lineage fields is implemented yet, so every engagement figure is timing-based. Table \ref{tab:stats} demonstrates the problem LARA addresses, whether the same metric means the same thing across interaction modes. §3.4 shows what LARA itself can measure today.

\subsection{Shift 1: the agent reads first }
\label{ssec:shiftone}

Assistant-wide CTR is near zero because 99.7\% of interactions never invoke retrieval, leaving no applicable denominator. Retrieval fired on 0.31\% of in-scope assistant prompts. Where retrieval fires, it works. The null rate is 1.51\% and users click about 1.05 times per returned grid. The direct test is ACR, which requires consumed\_asset\_id and is not yet computed. Externally, click-free consumption reached 68.0\% of U.S. search queries 
and rises when an AI-generated answer is shown \cite{ref11}.

\subsection{Shift 2: action is delayed }
\label{ssec:shifttwo}
Assistant sessions average 2.04 prompts (chat: 2.65), so retrievals are routinely followed by further turns in which the same asset can reappear. More directly, exports initiated from chat land in a different session\_guid once the user leaves the chat surface. Session-level attribution counts these users as abandoning, which inflates chat abandonment (59.5\%) and deflates engagement. Carrying source\_retrieval\_id on the export event removes this error.

\subsection{Shift 3: retrieval and generation blend (the reformulation reversal) }
\label{ssec:shiftthree}

A naive reading predicts that assistant users reformulate less than gallery users, since they do not inspect result lists. Chat users reformulate more: 44.4\% versus 33.1\%. The two numbers measure different behavior. Gallery reformulation means “these results were not relevant; I will try another query.” Chat reformulation mostly means “I see the generated output and want to change it” (“make it bluer,” “try a bolder font”), because most chat prompts never trigger retrieval. This is Shift 3 in production data: one behavioral signal mixes retrieval failure with generation refinement, and the aggregate cannot be read as either. LARA splits it into Search Reformulation, counted only on retrieval turns, and Downstream Correction, counted everywhere else. At the model level, the blending itself is well documented. Models often favor parametric knowledge over retrieved context when the two conflict \cite{ref12}.

\subsection{What LARA can measure today }

With current Express telemetry, null rate (1.51\%) and clicks per returned grid ($\approx$1.05) are computable from existing logs. Precision@K and NDCG@K need returned\_asset\_ids plus a labeled relevance sample; neither is computed yet. GPR, ACR, ORF, and UOR need consumed\_asset\_id, source\_retrieval\_id, and the retrieval role tag; today they require manual trace annotation. Retrieval-attributed export rate needs source\_retrieval\_id on export events, since exports cross sessions and are not computable from current logs. A team can confirm retrieval works when it fires, but not yet whether the user or the agent used what it returned. returned\_asset\_ids would unblock much of this list.

\section{Unexpected Finding: Retrieval Nearly Vanishes }
\label{sec:unexpected}

Gallery search is retrieval by definition; the assistant invoked retrieval on 0.31\% of prompts. Roughly 82\% of assistant activity is generative editing with no gallery-search analog. LARA’s layers apply once retrieval fires, covering only a small slice of interaction volume. The larger risk may be missed retrieval: the planner routes a prompt such as “find me a birthday invitation template” to generative editing, no retrieval event fires, and Layer 1 has nothing to score. Confirming this requires ground-truth intent. Table \ref{tab:stats}’s 10.7\% chat retrieval-intent figure is directional until reconciled against the 0.31\% fired-retrieval figure; if it reflects intent rather than fired retrieval, the gap between the two is itself a first estimate of missed retrieval.

We name this Layer 0, planner routing, and treat it as an extension of LARA rather than a fourth layer. This boundary is intentional. LARA addresses attribution once retrieval fires; the four-field schema is a concrete fix for that problem. Planner routing needs ground-truth intent labels this deployment does not have. Naming it here keeps the two problems separate without claiming either framework solves both.

Two measures follow. Mis-routing rate is the share of retrieval calls whose prompt falls outside retrieval scope; these calls should be excluded from Layer 1. Missed-retrieval rate is the share of prompts with retrieval intent that never triggered retrieval; it needs ground-truth intent compared against routing decisions, and no published standard provides one. Multi-intent prompts make routing harder still: in a 273-prompt sample, roughly 5–7\% contained two or more distinct retrieval intents. Instrumenting Layer 0 means logging the planner’s intent class and routing decision on every prompt, not only on prompts that reach retrieval. Today the planner’s generated query is captured only on the retrieval path, and raw prompt text on about 65\% of prompt-send events.

\section{Discussion and Conclusion }
\label{sec:disc}

Lessons for practitioners. Three lessons from Express apply beyond creative tools. First, do not report assistant-wide CTR or reformulation as retrieval metrics; restrict retrieval metrics to turns where retrieval fired, and split reformulation into Search Reformulation and Downstream Correction. Second, log the full ranked list at fire time before adding anything else; it is the only field that unblocks Layer 1. Third, define user-initiated use once and publish the definition. Express chat meaningful-action rate moved from 7.6\% to 40.5\% depending on which actions qualified.

Limitations. All evidence comes from one deployment, and the 0.31\% retrieval rate and the reformulation reversal may be specific to creative assistants dominated by generative editing. None of the lineage fields is implemented, so all Layer 2 figures are timing-based inferences; retrieval role is inferred post hoc rather than tagged at retrieval time; and chat abandonment is overstated by exports that land in separate sessions. LARA begins at the retrieval component; how the agent interprets intent, generates queries, or decomposes multi-part requests upstream needs separate instrumentation. Production telemetry is not inherently trustworthy \cite{ref5}; schema and value-range validation has not been run on this dataset; synthetic traffic was excluded manually; and latency and tokens per successful task are not instrumented per stage.


In sum, agentic assistants remove the three conditions behavioral retrieval metrics depend on, who sees the results, when the user acts, and what caused the outcome. LARA restores attribution with three independent layers and four logging fields, without new modeling. Adobe Express shows why this is needed. The same reformulation metric measures search failure on one surface and generation refinement on the other. It also shows where the next problem lies. When retrieval fires on 0.31\% of assistant prompts, the open question is not whether retrieval ranked well but whether it was called at all.

\section*{Presenter Biographies}

Yan (Fergie) Liang is a Product Manager at Adobe on the Search, Discovery, and Content AI (SDC) team, working on agentic retrieval, grounding, and evaluation for Adobe Public Content. She holds an MBA from Northwestern Kellogg.

Anuj Agarwal is a Data Scientist at Adobe with over ten years in applied data science, specializing in search and generative AI and how assistant search quality is measured from user behavior. He previously worked at Amazon on Prime Membership pricing and revenue optimization, and has authored research published by Gartner.

Aditya Reddy Cheruku is a Principal Data Scientist at Adobe with over ten years in applied data science, working on how search operates inside AI assistants and how its quality is measured from user behavior. He previously worked at Walmart on pricing and assortment, and is a co-inventor on Walmart-filed patents.

\section*{Company Description}

Adobe is a global technology company that provides creative, marketing, and document software used by individuals, teams, and enterprises worldwide. Its Creative Cloud portfolio includes Adobe Express, a web and mobile creative tool for fast, templated design work. The Search, Discovery, and Content AI (SDC) team builds the retrieval infrastructure that powers search and content discovery across Adobe Express, Stock, Templates, and related creative surfaces, including the agentic AI assistant studied in this paper.

\bibliographystyle{ACM-Reference-Format}
\bibliography{lara}

\end{document}